# Large magnetic anisotropy in quasi one-dimensional spin-½ fluoride antiferromagnets with a d(z$^2$)$^1$ ground state

D. Kurzydłowski,[1,2]* and W. Grochala[1]*

[1] *Centre of New Technologies, University of Warsaw, Banacha 2c, 02-097, Warsaw, Poland*

[2] *Faculty of Mathematics and Natural Sciences, Cardinal Stefan Wyszynski University in Warsaw, Wóycickiego 1/3, 01-938, Warsaw, Poland*

**Abstract** Hybrid density functional calculations are performed for a variety of systems containing $d^9$ ions ($Cu^{2+}$, $Ag^{2+}$), and exhibiting quasi-one-dimensional magnetic properties. In particular we study fluorides containing these ions in a rarely encountered compressed octahedral coordination which forces the unpaired electron in the $d(z^2)$ orbital. We predict that such systems should exhibit magnetic anisotropies surpassing that of $Sr_2CuO_3$ – one of the best realizations of an one-dimensional system known to date. In particular we predict that the inter-chain coupling in the $Ag^{2+}$-containing $[AgF][BF_4]$ should be nearly five orders of magnitude smaller than the intra-chain interaction. Our results indicate that quasi-one-dimensional spin-½ systems containing chains with spin sites in a $d(z^2)^1$ local ground state could constitute a versatile model for testing modern theories of quantum many-body physics in the solid state.

## I. INTRODUCTION

Coupling of structural, electronic, and magnetic degrees of freedom in compounds containing transition metals cations with partially filled *d* shells leads to many intriguing phenomena, such as charge-density waves,[1,2] or unconventional superconductivity.[3,4] Due to strong Coulomb interaction within the *d* shell such compounds are most often insulators,[5] with the unpaired spin density effectively localized on the transition metal cations. In such systems magnetic interactions between the unpaired spins, which are mediated via the superexchange mechanism, can be described by the Heisenberg Hamiltonian,

$$H = \sum_{i,j} J_{ij} \mathbf{S_i} \cdot \mathbf{S_j}, \qquad (1)$$

where $J_{ij}$ is the magnetic coupling constants between spin sites *i* and *j*. In this convention positive values of $J_{ij}$ correspond to antiferromagnetic (AFM) spin ordering, while negative values to a ferromagnetic (FM) one.

Much attention was paid to the study of magnetic interactions in compounds containing $Cu^{2+}$ cations ($3d^9$ electronic configuration) exhibiting quasi one-dimensional (1D) antiferromagnetic properties.[6,7] More recently, there has been an upsurge of interest in homologous systems featuring $Ag^{2+}$ cations with a $4d^9$ electronic configuration.[8–10] Both kinds of systems are composed of spin-½ Heisenberg chains characterized by an intra-chain coupling constant, $J_{1D}$, which can be defined by taking the Heisenberg Hamiltonian in the form given by equation (1) and considering only nearest-neighbors along the chain:

$$H = J_{1D} \sum_i \boldsymbol{S_i} \cdot \boldsymbol{S_{i+1}}, \qquad (2)$$

The ground state of an 1D AFM system composed of isolated chains is disordered (Luttinger liquid).[11,12] This state is quantum critical, with very small inter-chain interactions leading to three-dimensional magnetic ordering at finite Néel temperatures ($T_N$).[13–15] However even in the ordered state the magnetic moments are extremely small due to quantum fluctuations.[16] Although 1D AFM systems have been intensely studied for more than half a century, new physical phenomena, such as unusual magnetic excitations,[17] are still being discovered.

Theoretical models of one-dimensional spin systems, such as those describing the spin transport mechanism,[18,19] are best tested on systems exhibiting a very small ratio between the Néel temperature and the intra-chain coupling constant. One of the best examples of such systems is $Sr_2CuO_3$ which exhibits $T_N$ equal to 5.4 K,[16] and $J_{1D} = 2785$ K,[11,20] thus yielding a $T_N/J_{1D}$ ratio of $1.9 \cdot 10^{-3}$. Recently we performed calculations of the intra-chain coupling constant for a handful of 1D AFM systems, including $Sr_2CuO_3$.[8] We predicted that that [AgF][BF$_4$], a compound containing $Ag^{2+}$ cations, should exhibit $J_{1D}$ equal to ~3840 K, thus possibly surpassing in strength the AFM interactions exhibited by $Sr_2CuO_3$, and making it a good candidate for a model 1D AFM system, provided that the inter-chain interactions are weak.

Following these findings we present here calculations of relatively weak inter-chain coupling constants ($J_\perp$) for a variety of $Cu^{2+}$ and $Ag^{2+}$ compounds exhibiting quasi one-dimensional properties. By including some experimentally well-studied systems ($KCuF_3$, $Sr_2CuO_3$, $Ca_2CuO_3$, $KAgF_3$) we verify that our method is capable of reproducing the measured $T_N/J_{1D}$ and $|J_\perp|/J_{1D}$ ratioa with good accuracy. We confirm the 1D nature of [AgF][BF$_4$] by predicting $T_N/J_{1D}$ equal to $5.5 \cdot 10^{-4}$, and single-out other compounds that exhibit nearly ideal 1D AFM properties (*i.e.* low $T_N/J_{1D}$ ratios and high magnetic anisotropy). In particular we find that several compounds containing $Cu^{2+}/Ag^{2+}$ cations in rare a compressed octahedral geometry, which enforces single-occupation of the $d(z^2)$ orbital, exhibit magnetic anisotropies surpassing that of the presently best 1D AFM systems (all of which containing the unpaired electron in a $d(x^2-y^2)$ orbital).

## II. COMPUTATIONAL DETAILS

For each of the studied compounds we consider the nearest-neighbor interactions within the AFM-coupled chain ($J_{1D}$), as well as diverse inter-chain interactions. Depending on the structure of the compound there might be more than one type of inter-chain couplings, in this case we will label those $J_\perp^1, J_\perp^2$, etc. For each type of the superexchange coupling topology we present a complete analysis of all relevant superexchange interactions, and derive models of magnetic states appropriate for the extraction of the inter-chain coupling constants.

Due to the single-determinant nature of DFT calculations we obtained the values of the coupling constants with the use of the 'broken symmetry' method,[21] which makes use of the Ising Hamiltonian

$$H = \sum_{i,j} J_{ij} \boldsymbol{S_i^z} \cdot \boldsymbol{S_j^z}, \qquad (3)$$

with this method one can derive the values of the values of $J_{ij}$ from the energies of different spin states (vide infra). It was shown that the coupling constants extracted from the Ising Hamiltonian can indeed correspond to those of the Heiseberg Hamiltonian.[22]

Solid state collinear calculations were performed with the HSE06 functional[23] which is a hybrid functional mixing the Density Functional Theory Generalized Gradient Approximation (DFT-GGA) functional of Perdew, Burke, and Ernzerhof,[24] with 25% of Hartree-Fock (HF) exchange energy. We found out previously that use of the HSE06 functional has led to a good reproduction of the intra-chain superexchange coupling constants with a systematic and rather small 11% overestimation of their values.[8] The projector-augmented-wave method[25,26] was used as implemented in the VASP 5.2 code.[27–30] Valence electrons were treated explicitly, while standard VASP pseudopotentials (accounting for scalar relativistic effects) were used for the description of core electrons. We used a plane-wave basis set with a cut-off energy of 920 eV which was lowered to 850 eV for Cs- and Sr-containing compounds (Cs, and Sr pseudopotentials do not allow for higher cut-off energies). The energy convergence criterion was $2 \cdot 10^{-7}$ eV (= $2 \cdot 10^{-4}$ meV) per formula unit (f.u.). We used a fine $k$-point mesh with a spacing of $0.03 \cdot 2\pi$ Å$^{-1}$. The band energy was smeared with the used of the tetrahedron method with Blöchl corrections.

In our calculations we do not include non-collinear magnetic interactions. Such interactions are most often a result of spin-orbit coupling, which is small for $d^9$ ions in octahedral coordination due to orbital momentum quenching. In this work we focus on the relative strength of the inter- and intra-chain interaction, which is mainly influenced by the strength of the superexchange interactions and not secondary spin-orbit coupling effects.

Before performing single point calculations to determine the values of the magnetic coupling constants we optimized the geometry of every system by performing full relaxation of both the unit cell parameters and of the atomic coordinates. This optimization was conducted with the HSE06 functional and the parameters as given above. The convergence criterion for the relaxation was: forces below 0.015 eV/Å, and pressure below 1 kbar. The geometry optimization was conducted for the spin state of lowest energy (for a detailed comparison of the calculated and experimental geometries see ref. 8). We verified that the parameters employed in the calculations ensured convergence of the superexchange constants to within $2\cdot10^{-3}$ meV (0.02 K). Visualization of structures and volumetric data has been performed with the use of the VESTA software.[31]

## III. RESULTS AND DISCUSSION

We have performed calculations for fifteen compounds containing $Cu^{2+}$ and $Ag^{2+}$ cations. These distinct quasi-1D systems can be grouped accordingly to the structural features of the AFM-coupled chains. Due to the operation of the Jahn-Teller (JT) effect $MX_6$ complexes (M – $d^9$ cation, X – ligand) can exhibit either compressed or elongated octahedral geometry ($D_{4h}$ symmetry). In the former scenario the underlying electronic state of metal cation corresponds to half-occupation of the $d(z^2)$ orbital of $A_{1g}$ symmetry, while elongation forces the unpaired electron into the $d(x^2-y^2)$ orbital with $B_{1g}$ symmetry (in both cases we define the $z$ axis as parallel to the axis of the JT distortion).

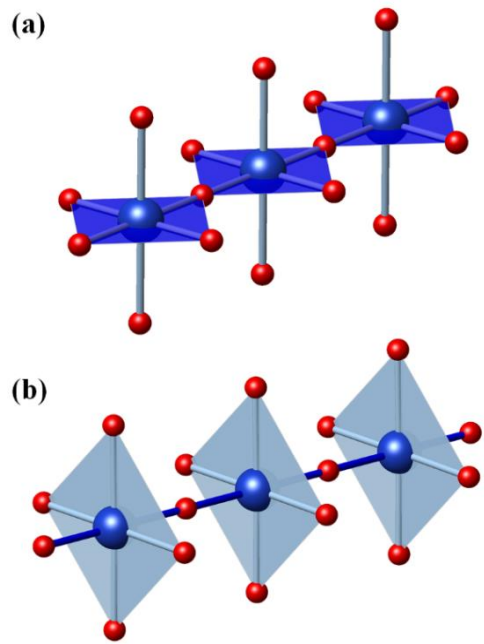

FIG. 1 (Color online) (a) Arrangement of elongated $MX_6$ octahedral leading to formation of chains featuring half-occupation of the $B_{1g}$ orbital of the $M^{2+}$ cation. (b) Arrangement of compressed $MX_6$ octahedral leading to formation of chains featuring with half-occupied $A_{1g}$ cation orbitals. Blue/red atoms balls depict M/X atoms, long/short M-X bonds are depicted with light blue/dark blue color.

Based on the Goodenough-Anderson-Kanamori (GKA) rules[32–34] strong AFM coupling can be expected for chains composed of either: (i) elongated $MX_6$ octahedra sharing equatorial (shorter) bonds, or (ii) compressed $MX_6$ octahedra sharing axial (shorter) bonds. Throughout this work we will describe the former structure motif, depicted in Fig. 1(a), as "$B_{1g}$ chains", and the latter [Fig. 1(b)] as "$A_{1g}$ chains". Note that for $A_{1g}$ chains the JT axis is parallel to the direction of the chain propagation, while for $B_{1g}$ chains the JT distortion takes place in the perpendicular direction.

### A. Fluorides with $B_{1g}$ chains

We start our discussion with fluorides containing $Cu^{2+}$ and $Ag^{2+}$ cations with a general formula $M'MF_3$ (M' = Na, K, Rb, Cs, Ag; M = Cu, Ag). All of these compounds adopt structures that can be derived from the perovskite polytype. One of the most studied member of this family is $KCuF_3$, which is a prototypical 1D AFM system.[35–43]

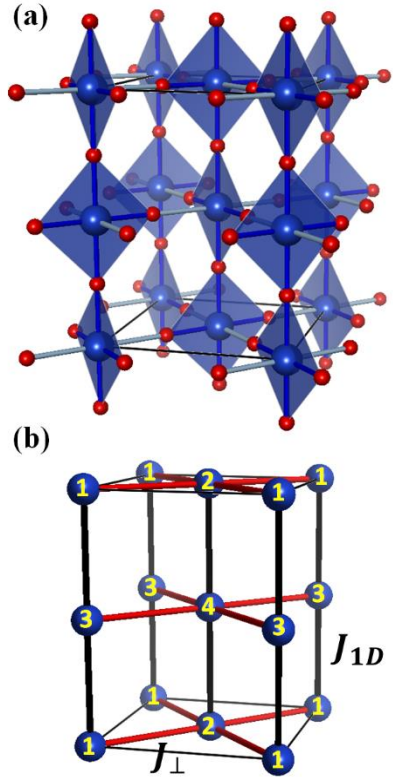

FIG. 2 (Color online) (a) The structure of a-$KCuF_3$; K atoms were omitted for clarity; (b) Schematic description of the connectivity of the $J_{1D}$ (black) and $J_\perp$ (red) superexchange paths.

This compound exhibits two polymorphs: a-$KCuF_3$ of *I4/mcm* symmetry, and d-$KCuF_3$ belonging to the *P4/mbm* space group.[36] Due to orbital ordering first described by Kugel and Khomskii,[44] both structures consist of $B_{1g}$-type chains separated by $K^+$ cations. The difference between the two polytypes lies in the relative orientation of $CuF_4$ plaquettes neighboring along the chain. As this feature does not markedly influence the superexchange interactions we have conducted calculations for the larger unit cell polytype ($Z = 4$), a-KCuF$_3$, depicted in FIG. 2(a).

Due to the high symmetry of KCuF$_3$ there exists only one relevant inter-chain interaction ($J_\perp$) shown in FIG. 2(b). Based on GKA rules it is expected that inter-chain coupling will be ferromagnetic, that is $J_\perp < 0$. Indeed the experimental value of $J_\perp$ is –21 K,[45] while the intra-chain coupling constant ($J_{1D}$) is equal to 406 K.[37] The non-zero inter-chain interaction leads to a finite Néel temperature of 39 K, hence $T_N/J_{1D} = 9.6 \cdot 10^{-2}$.

The value of the Néel temperature is dependent on both the strength of the inter- and intra-chain coupling. Recent quantum Monte Carlo simulations on 1D AFM systems enabled linking the value of the inter-chain interaction with $T_N$ and $J_{1D}$ for a 1D system:

$$|J_\perp| = T_N / \left[ 4c \sqrt{ \ln\left(\frac{\lambda J_{1D}}{T_N}\right) + \frac{1}{2} \ln\ln\left(\frac{\lambda J_{1D}}{T_N}\right) } \right], \qquad (4)$$

where $c = 0.23$, and $\lambda = 2.6$.[15] Inserting the experimental values of $T_N$ and $J_{1D}$ determined for KCuF$_3$ into Eq. (4) yields $|J_\perp|$ equal to 20.9 K, in very good agreement with the experimental value.

TABLE 1. Magnetic states of KCuF$_3$. Spin up/down sites are indicated with a +/– sign. Site labelling follows that of FIG. 2(b), $E_{nm}$ denotes the part of the total-energy of the system which is independent of the spin state.

| State | Site 1 | 2 | 3 | 4 | Energy per f.u. |
|---|---|---|---|---|---|
| F1 | + | + | + | + | $0.25 J_{1D} + 0.5 J_\perp + E_{nm}$ |
| A1 | + | + | – | – | $-0.25 J_{1D} + 0.5 J_\perp + E_{nm}$ |
| A2 | + | – | + | – | $-0.25 J_{1D} - 0.5 J_\perp + E_{nm}$ |

In order to extract magnetic coupling constants with the use of the broken-symmetry method we have constructed three spin states of a-KCuF$_3$: (i) F1 state with intra- and inter-chain FM coupling, (ii) A1 state with AFM coupling within chains and FM coupling between them, and (iii) A2 state with AFM coupling along both superexchange routes. Employing the Ising Hamiltonian, Eq. (3), enables calculation of the energy of each state with respect to $J_{1D}$ and $J_\perp$. The corresponding formulas are summarized in Table 1 together with the direction of the unpaired spins on each Cu$^{2+}$ site of the KCuF$_3$ crystal structure [Fig. 2(b)]. The formulas from Table 1 can be combined in order to extract the superexchange constants

$$J_{1D} = 2F1 - 2A1, \qquad (5)$$

$$J_\perp = A1 - A2. \qquad (6)$$

By calculating the energies of the F1, A1, and A2 magnetic states with the use of the HSE06 hybrid functional we obtain the $J_{1D}$ and $J_\perp$ values of 594 K, and –48 K, respectively. These values compare to those obtained in previous periodic calculations of Moreira *et al.* which employed the B3LYP hybrid functional ($J_{1D} = 652$ K, $J_\perp = -23.2$ K).[46]

With the theoretical values of $J_{1D}$ and $J_\perp$ we can obtain the value of $T_N$ with the use of Eq. (4), and hence calculate the $T_N/J_{1D}$ ratio. For KCuF$_3$ we obtain in our calculations $T_N/J_{1D} = 14 \cdot 10^{-2}$, which is rather close to the experimental value of $9.6 \cdot 10^{-2}$ (taking theoretical values reported in ref. 47 one arrives at $T_N/J_{1D} = 3.6 \cdot 10^{-2}$). The overestimation of the experimental value in our calculations can be traced back to the underestimation of the magnetic anisotropy, as the theoretical $|J_\perp|/J_{1D}$ ratio is ($8.1 \cdot 10^{-2}$) is smaller than the experimental one ($5.2 \cdot 10^{-2}$).

For other perovskite fluorides of copper and silver the inter- and intra-chain couplings can be extracted by the same procedure as for KCuF$_3$. Apart from CsAgF$_3$ which is iso-structural with a-KCuF$_3$,[47] the structure of these compounds (KAgF$_3$ – ref. 9; RbAgF$_3$ – ref. 47; NaCuF$_3$/AgCuF$_3$ – ref. 48) can be derived from that of a-KCuF$_3$ by introduction of tilting of the MX$_6$ octahedra.[49] This distortion however does not alter the topology of the superexchange pathways, and therefore the spin states given in Table 1 can be used to extract $J_{1D}$ and $J_\perp$ through Eq. (5) and Eq. (6). In Table 2 we report the calculated values of $J_\perp$ together with the values of the $|J_\perp|/J_{1D}$ and $T_N/J_{1D}$.

TABLE 2. Values of the magnetic coupling constants, and the $|J_\perp|/|J_{1D}|$, $T_N/J_{1D}$ ratios calculated for fluorides of divalent copper and silver exhibiting B$_{1g}$ chains. The values of $J_{1D}$ are taken from ref. 8. Theoretical $T_N$ values were calculated from $J_\perp$ and $J_{1D}$ using Eq. (4). Experimental data, in parenthesis, for KCuF$_3$ (ref. 38, 46), AgCuF$_3$/NaCuF$_3$ (ref. 49), and KAgF$_3$ (ref. 9) are given for comparison.

|  | $J_{1D}$ (K) | $J_\perp$ (K) | $|J_\perp|/J_{1D}$ ($10^{-3}$) | $T_N/J_{1D}$ ($10^{-3}$) |
|---|---|---|---|---|
| KCuF$_3$ | 594 (406) | –48 (–21) | 81 (52) | 140 (96) |
| AgCuF$_3$ | 436 (298) | –61 | 140 | 222 (67) |
| NaAgF$_3$ | 369 (191) | –59 | 160 | 249 (94) |
| CsAgF$_3$ | 1867 | –106 | 57 | 103 |
| RbAgF$_3$ | 1669 | –83 | 49 | 92 |
| KAgF$_3$ | 1311 (1160) | –58 | 44 | 82 (57) |

Although magnetic susceptibility of CsAgF$_3$ and RbAgF$_3$ was measured,[47] the values of $T_N$ and $J_{1D}$ were not determined for these systems. Our calculations agree well with the experimental data of KAgF$_3$ with $T_N/J_{1D}$ slightly overestimated compared to the experimental one. The largest discrepancy between theory and experiment is found for AgCuF$_3$ and NaAgF$_3$. Taking the experimental ordering temperatures and intra-chain coupling constants of these systems,[48] one arrives at $T_N/J_{1D}$ ratios considerably lower than the theoretical ones. However, the magnetic ordering temperatures of AgCuF$_3$ and NaCuF$_3$ were derived from susceptibility measurements of samples containing FM impurities. The authors took $T_N$ as the temperature corresponding to a peak in the magnetic susceptibility, but this peak is found in the region where the signal from FM impurities is dominant. It thus possible that long range magnetic order sets out at higher temperatures than those initially presumed,[48] but remains masked by the Curie-like temperature dependence of the impurities. In this case the experimental $T_N/J_{1D}$ values would be higher than reported, and thus closer to our theoretical values.

Comparing $T_N/J_{1D}$ ratios calculated for silver(II) fluorides and copper(II) fluorides indicates that the former compounds exhibit larger magnetic anisotropies, despite being characterized by stronger inter-chain FM coupling. This is mostly a consequence of the much stronger intra-chain AFM interactions. This feature, in turn, originates from a much more pronounced hybridization of the Ag(*4d*) states with F(*2p*) than that taking place for the Cu(*3d*) and F(*2p*) states.[50]

### B. Oxides with B$_{1g}$ chains

We now turn to two oxides systems containing Cu$^{2+}$ cations in an elongated octahedral coordination, Sr$_2$CuO$_3$ and Ca$_2$CuO$_3$. Both compounds are considered as one of the best realizations of a quasi-1D AFM-coupled systems among all inorganic compounds, and they constitute important references in our quest for model 1D AFM systems.[51,52] Their structure, depicted in FIG. 3(a), consists of B$_{1g}$-type chains with linear Cu-O-Cu bidges.[53,54]

We distinguish two intra-chain superexchange paths, which are depicted in FIG. 3(b). The shorter one is $J_\perp^1$, and it links the AFM-coupled chains into 2D sheets. Given the fact that ideal 2D systems do not display long-range order at finite temperatures, one must also include another superexchange path that couples the 2D sheets. Therefore a longer route, characterized by the $J_\perp^2$ coupling constant, is also taken into account in our model. Both $J_\perp^1$ and $J_\perp^2$ can be combined into an effective intra-chain coupling ($J_\perp^{eff}$)

$$\left|J_\perp^{eff}\right| = \frac{\sum_n z_n |J_\perp^n|}{\sum_n z_n}, \quad (7)$$

where $z_n$ is the number of neighbors along a given inter-chain exchange coupling route (for B$_2$CuO$_3$ $z_1 = 4$, $z_2 = 8$). The effective inter-chain interaction can be used with the value of $J_{1D}$ to predict the value of $T_N$ through Eq. (4).

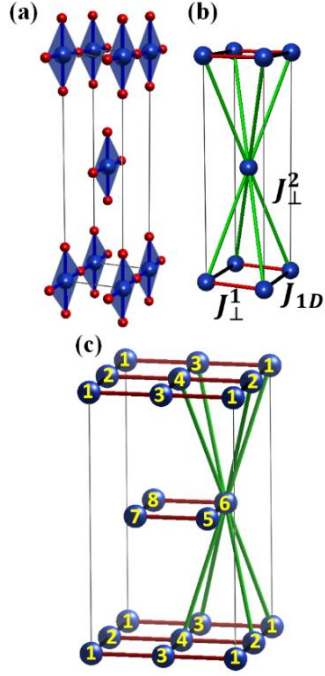

FIG. 3 (Color online) (a) The structure of M'$_2$CuO$_3$ (M' = Ca, Sr); B atoms were omitted for clarity; (b) Schematic description of the connectivity of the $J_{1D}$ (black line), $J_\perp^1$ (red line), and $J_\perp^2$ (green line) superexchange paths; (c) The 1x2x2 supercell used for the calculation of coupling constants.

TABLE 3. Magnetic states of M'$_2$CuO$_3$ (M' = Sr, Ca) . Spin up/down sites are indicated with a +/– sign; their labelling follows that of FIG. 3(c).

| State | Site | | | | | | | | Energy per f.u. |
|---|---|---|---|---|---|---|---|---|---|
| | 1 | 2 | 3 | 4 | 5 | 6 | 7 | 8 | |
| F1 | + | + | + | + | + | + | + | + | $0.25J_{1D} + 0.25J_\perp^1 + J_\perp^2 + E_{nm}$ |
| A1 | + | − | + | − | + | − | + | − | $-0.25J_{1D} + 0.25J_\perp^1 + E_{nm}$ |
| A2 | + | + | − | − | + | + | − | − | $0.25J_{1D} - 0.25J_\perp^1 + E_{nm}$ |
| A3 | + | + | + | + | − | − | − | − | $0.25J_{1D} + 0.25J_\perp^1 - J_\perp^2 + E_{nm}$ |

In order to calculate the three magnetic coupling constants ($J_{1D}$, $J_\perp^1$, and $J_\perp^2$) four spin states have to be taken into account – these are described in Table 3. Apart from F1 all of these states require extending the structure of B$_2$CuO$_3$ into a 1x1x2 supercell, and therefore all calculations were performed in this supercell, which is depicted in FIG. 3(c).

Equations (8) – (9) give the values of the coupling constants with respect to the magnetic states

$$J_{1D} = F1 - 2A1 + A3, \qquad (8)$$

$$J_\perp^1 = F1 - 2A2 + A3, \qquad (9)$$

$$J_\perp^2 = 0.5F1 - 0.5A3. \qquad (10)$$

The values of the coupling constants are summarized in Table 4. The interaction along the $J_\perp^1$ superexchange route is weakly antiferromagnetic, in accordance with previous calculations of de Graaf and Illas.[55] This AFM coupling is most probably a result of the direct interaction between $d(x^2-y^2)$ orbitals of neighboring chains. This notion is further corroborated by the fact that $J_\perp^1$ is larger for $Ca_2CuO_3$, which exhibits shorter inter-chain separations than its Sr analogue.[55] Analogous weak AFM interactions are observed in $CuCl_2 \cdot 2H_2O$,[56] as well as $Na_2AgF_4$,[10] both of which exhibit similar stacking of plaquettes containing spin-½ cations.

TABLE 4. Values of the magnetic coupling constants, and the $|J_\perp^{eff}|/J_{1D}$, $T_N/J_{1D}$ ratios calculated for $B_2CuO_3$. The values of $J_{1D}$ are taken from ref. 8. Theoretical $T_N$ values were calculated from $J_\perp$ and $J_{1D}$ using Eq. (4). The experimental $|J_\perp^{eff}|/J_{1D}$ ratio (given in parenthesis) was estimated from experimental $J_{1D}$ and $T_N$ values ($Sr_2CuO_3$ – ref. 16; $Ca_2CuO_3$ – ref. 51) using Eq. (4). In case of $Ca_2CuO_3$ we assumed $J_{1D}$ = 2680 K as derived from rigourous configuration-interaction calculations (ref. 55).

|  | $J_{1D}$ (K) | $J_\perp^1$ (K) | $J_\perp^2$ (K) | $|J_\perp^{eff}|$ (K) | $|J_\perp^{eff}|/J_{1D}$ ($10^{-3}$) | $T_N/J_{1D}$ ($10^{-3}$) |
|---|---|---|---|---|---|---|
| $Sr_2CuO_3$ | 3058 (2797) | 17 | –0.4 | 3.8 | 1.3 (0.7) | 3.2 (1.9) |
| $Ca_2CuO_3$ | 2961 | 24 | 2.1 | 6.5 | 2.2 (1.6) | 5.4 (4.1) |

As in the case of $KCuF_3$ and $KAgF_3$ we observe quite good accordance between our calculations and the experimental values of $|J_\perp^{eff}|/J_{1D}$ and $T_N/J_{1D}$, with both ratios slightly overestimated in the calculations. Unfortunately there are no oxide systems of divalent silver with which $B_2CuO_3$ compounds can be compared. This is a result of the strong tendency towards disproportionation of $Ag^{2+}$ to diamagnetic $Ag^+$ and low-spin $Ag^{3+}$ in the oxide environment, which cannot be reversed even under application of high pressure.[57]

### C. Fluorides with $A_{1g}$ chains

Although an axial elongation is more commonly found than compression for $d^9$ systems with deformed octahedral coordination,[58–62] there are several systems containing $Cu^{2+}$ and $Ag^{2+}$ cations in the genuine local $A_{1g}$ ground state (*i.e.* with half-occupation of the local $d(z^2)$ orbital of the metal). Among these $[MF][AsF_6]$ (M = Cu, Ag), M'$CuAlF_5$ (M' = K, Cs), and $CsAgAlF_6$ adopt structures containing $A_{1g}$ chains with bent M-F-M bridges (M = Cu, Ag).[62–67] As shown in the Supplemental Material (ref. 68), these systems exhibit the same topology of the superexchange paths as $KCuF_3$. Consequently spin states described in Table 1 can be used for the extraction of the intra- and inter-chain coupling via Eq. (5) and (6). The results of the calculations are summarized in Table 6.

TABLE 5. Values of the magnetic coupling constants, and the $|J_\perp|/|J_{1D}|$, $T_N/J_{1D}$ ratios calculated for fluorides of divalent copper and silver exhibiting $A_{1g}$ chains. The values of $J_{1D}$ are taken from ref. 8. Theoretical $T_N$ values were calculated from $J_\perp$ and $J_{1D}$ using Eq. (4).

|  | $J_{1D}$ (K) | $J_\perp$ (K) | $|J_\perp|/J_{1D}$ ($10^{-3}$) | $T_N/J_{1D}$ ($10^{-3}$) |
| --- | --- | --- | --- | --- |
| [CuF][AsF$_6$] | 404 | 0.5 | 1.3 | 3.4 |
| CsCuAlF$_6$ | 292 | 0.1 | 0.3 | 0.8 |
| KCuAlF$_6$ | 162 | −0.1 | 0.8 | 2.2 |
| [AgF][AsF$_6$] | 1418 | −5.1 | 3.6 | 8.5 |
| CsAgAlF$_6$ | 670 | −0.6 | 0.9 | 2.5 |

The magnetic ordering temperatures for the studied $A_{1g}$ fluorides were not determined (although for $Cu^{2+}$-containing compounds they lie below 6K – see ref. 66), and therefore no comparison can be made between our results and the experimental ones. It is noteworthy to point, however, to the extremely small value of $T_N/J_{1D}$ (comparable or even lower than that of M'$_2$CuO$_3$) found for this group of compounds.

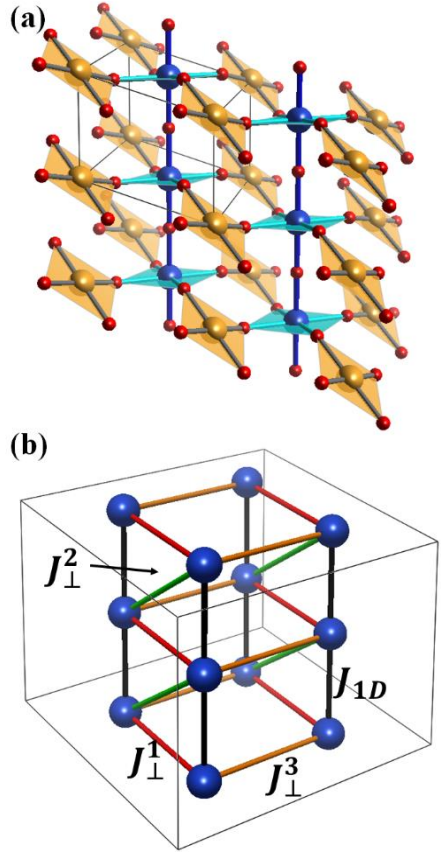

FIG. 4 (Color online) (a) Structure of [CuF][AuF$_4$] with gold atoms marked as orange balls; (b) the 2x2x2 supercell used for calculations of magnetic coupling constants, together with the depiction of the super-exchange paths.

Apart from the compounds mentioned above there are two more important and structurally characterized fluoride systems exhibiting $A_{1g}$ chains, namely [CuF][AuF$_4$],[69] and [AgF][BF$_4$],[65] which both contain nearly perfectly linear M-F-M bridges. Both are magnetically dense systems, with many relatively short secondary M⋯M separations, and a complex topology, and therefore the correct description of the interactions between the AFM chains requires taking into account more than one inter-chain coupling pathway.

At ambient conditions [CuF][AuF$_4$] crystallizes in a monoclinic cell containing $A_{1g}$ chains separated by AuF$_4^-$ units [FIG. 4(a)]. To account for all of the shortest inter-chain interactions we include in our model three inter-chain couplings, as shown in FIG. 4(b). In order to extract the J-values at least five spin-states have to be constructed within a 2x2x2 supercell.[68]

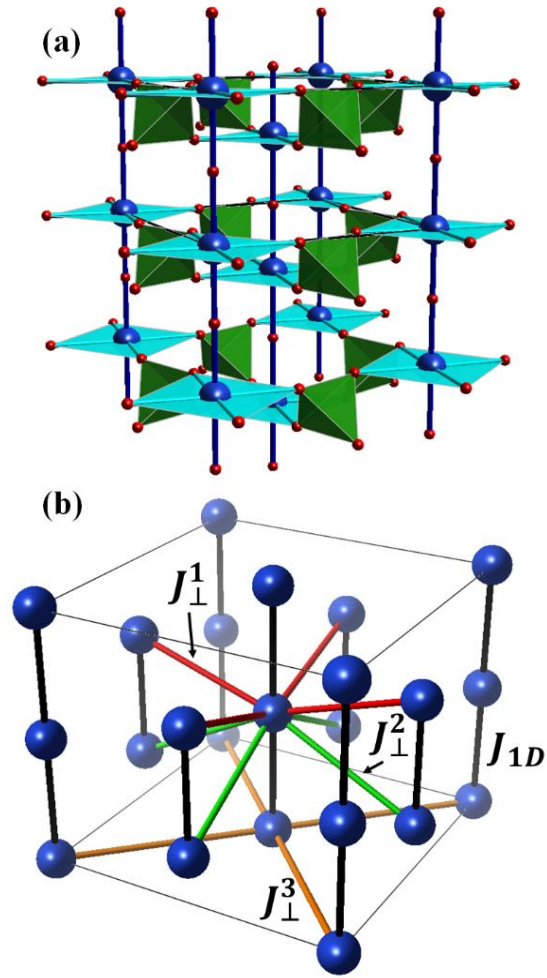

FIG. 5 (Color online) (a) Structure of [AgF][BF4] with BF$_4^-$ anions marked as green tetrahedra; (b) The √2x√2x2 supercell used for calculations of magnetic coupling constants, together with the depiction of the super-exchange paths.

The tetragonal structure of [AgF][BF$_4$] can be described as containing $A_{1g}$ chains, with a linear and nearly symmetric Ag-F-Ag bridges linked by BF$_4^-$ tetrahedra [FIG. 5(a)]. Apart from the intra-chain coupling constant there are three inter-chain exchange routes [FIG. 5(b)]. The spin states and their relation to the coupling constant values are given in ref. 68.

TABLE 6. Values of the inter-chain magnetic coupling constants, and the $|J_\perp^{eff}|/J_{1D}$, $T_N/J_{1D}$ ratios calculated for [CuF][AuF$_4$] and [AgF][BF$_4$] ($J_{1D}$ from ref. 8). Theoretical $T_N$ values were calculated from $J_\perp$ and $J_{1D}$) using Eq. (4).

|  | $J_{1D}$ (K) | $J_\perp^1$ (K) | $J_\perp^2$ (K) | $J_\perp^3$ (K) | $|J_\perp^{eff}|$ (K) | $|J_\perp^{eff}|/J_{1D}$ ($10^{-3}$) | $T_N/J_{1D}$ ($10^{-3}$) |
|---|---|---|---|---|---|---|---|
| [CuF][AuF$_4$] | 862 | −0.3 | 2.1 | −2.7 | 1.7 | 2 | 4.9 |
| [AgF][BF$_4$] | 3843 | −0.4 | −0.1 | 1.6 | 0.7 | 0.2 | 0.6 |

For both [CuF][AuF$_4$], and [AgF][BF$_4$] an effective inter-chain interactions can be defined using Eq. (7), with $z_1 = z_2 = z_3 = 2$ for the former compound, and $z_1 = z_2 = z_3 = 4$ for the latter. As in the case of B$_2$CuO$_3$ we use the value of $|J_\perp^{eff}|$ together with $J_{1D}$ to obtain the magnetic ordering temperature, and consequently $T_N/J_{1D}$. The results, summarized in Table 6, indicate that although both compounds are characterized by inter-chain interactions of similar strength, [AgF][BF$_4$] is a better realization of an 1D AFM system than [CuF][AuF$_4$], which is a result of a stronger intra-chain coupling for the former system.[8]

### D. Next-nearest-neighbor interactions within chains

For strongly AFM-coupled systems interactions beyond nearest-neighbors might be significant, as exemplified by the ring exchange observed in the 2D AFM systems La$_2$CuO$_4$.[70] In case of 1D AFM systems exhibiting linear M-F-M and M-O-M bridges we therefore considered next-nearest-neighbor (NNN) coupling ($J_{1D}^{NNN}$) along the chains.

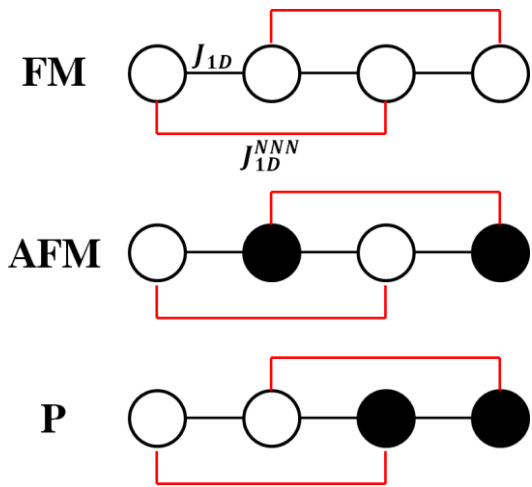

FIG. 6 (Color online) Depiction of the magnetic states used for calculations of $J_{1D}^{NNN}$, open/filled balls depict spin up/down sites.

In order to extract the value of $J_{1D}^{NNN}$ a magnetic states with paired spins (P state in FIG. 6) has to be taken into account. For this state the contributions from nearest-neighbor interactions cancel out, while the contribution from the NNN coupling is $-1/4 J_{1D}^{NNN}$ per spin site. For chains characterized by both FM and AFM order taking the NNN interaction into account shifts the energy by $1/4 J_{1D}^{NNN}$ per spin site. Consequently the value of $J_{1D}^{NNN}$ can be calculated through the following expression:

$$J_{1D}^{NNN} = F\text{M} + A\text{FM} - 2P, \qquad (11)$$

The values of the NNN coupling, summarized in Table 7, are sizable for systems exhibiting strong nearest-neighbor interaction ([AgF][BF$_4$], CsAgF$_3$, Sr$_2$CuO$_3$). The ratio between $J_{1D}^{NNN}$ and $J_{1D}$ does not exceed $6 \cdot 10^{-2}$, similarly to La$_2$CuO$_4$ for which the ratio between NN and NNN interactions within the [CuO$_2$] sheets was determined experimentally to of an order of $1.5 \cdot 10^{-2}$.[70] The predicted positive sign of both $J_{1D}^{NNN}$ and $J_{1D}$ hints at a possible weak spin frustration within the 1D chains featured both in [AgF][BF$_4$] and Sr$_2$CuO$_3$, due to the second-nearest neighbor intra-chain coupling.

TABLE 7. Values of next-nearest-neighbor intra-chain interactions ($J_{1D}^{NNN}$) calculated for selected systems.

|  | $J_{1D}^{NNN}$ (K) | $|J_{1D}^{NNN}|/|J_{1D}|$ ($10^{-2}$) |
|---|---|---|
| [AgF][BF$_4$] | 228 | 5.9 |
| CsAgF$_3$ | 54 | 2.9 |
| Sr$_2$CuO$_3$ | 148 | 4.8 |
| [CuF][AuF$_4$] | 7 | 0.9 |
| KCuF$_3$ | 0.2 | 0.04 |

### E. Summary

In order to assess the accuracy of the HSE06 functional for prediction of inherently weak inter-chain superexchange, we compared the experimentally known $T_N/J_{1D}$ ratios with those calculated here. As follows from FIG. 7 there exists a linear relationship between theoretical and experimental results with the former being usually overestimated by 45.3±0.4 %. The error is larger for AgCuF$_3$ and NaCuF$_3$, for which, however, there are doubts about the true value of their magnetic ordering temperatures as derived from experimental data (see Section A).

A summary of the calculated $T_N/J_{1D}$ and $|J_\perp^{eff}|/J_{1D}$ ratios, given in FIG. 8, shows that among the studied compounds four exhibit magnetic anisotropies larger than that of Sr$_2$CuO$_3$. Moreover all of these systems exhibit $A_{1g}$-type chains containing either Cu$^{2+}$ (CsCuAlF$_6$, KCuAlF$_6$), or Ag$^{2+}$ cations ([AgF][BF$_4$], CsAgAlF$_6$).

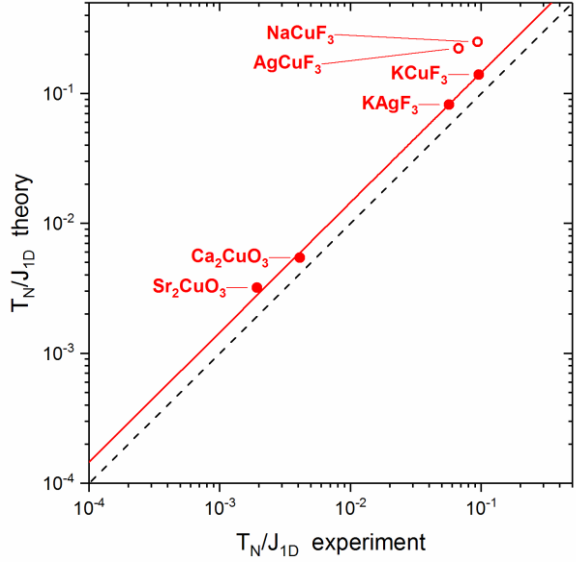

FIG. 7 Comparison of experimental and theoretical values of the $T_N/J_{1D}$ ratios for compounds which feature the unpaired electron in the $d(x^2-y^2)$ orbital of a metal. The red line marks a linear regression passing through (0, 0) with the slope of 1.453 ± 0.004. The values for $AgCuF_3$, and $NaCuF_3$, not included into the fit, are shown by open symbols.

In particular [AgF][BF$_4$] is characterized by a calculated $|J_\perp^{eff}|/J_{1D}$ ratio of only $1.9 \cdot 10^{-4}$, and the resulting computed $T_N/J_{1D}$ ratio is equal to $5.5 \cdot 10^{-4}$. These values are close to those estimated from experiment for DEOCC-TCNQF$_4$ ($0.2 \cdot 10^{-4}$ and $0.6 \cdot 10^{-4}$, respectively for $|J_\perp^{eff}|/J_{1D}$ and the scaled Néel temperature), an organic radical-ion salt claimed as the best known realization of an 1D AFM system.[45] Indeed, the moderate value of the intra-chain coupling constant of DEOCC-TCNQF$_4$ (110 K) results in a very low Néel temperature, with 3D magnetic order not detectable even at 20 mK.[45] In contrast we predict that [AgF][BF$_4$] should exhibit a magnetic ordering temperature of about 2 K. We note that our calculations do not take into account non-collinear interaction (e.g. the Dzyaloshinsky-Moriya coupling), as well as possible frustration of inter-chain interactions, both of which can significantly influence the value of the magnetic ordering temperature for this compound (as well as for the M'$_2$CuO$_3$ reference systems).[71,72] However more importantly the extremely strong AFM intra-chain interaction found for [AgF][BF$_4$] compound ($J_{1D}$ exceeding 3000 K) should enable testing the spin dynamics of this system even at temperatures above the magnetic ordering, similarly to the case of Sr$_2$CuO$_3$. In view of these results, experimental examination of magnetic features of [AgF][BF$_4$] certainly constitutes an interesting goal.

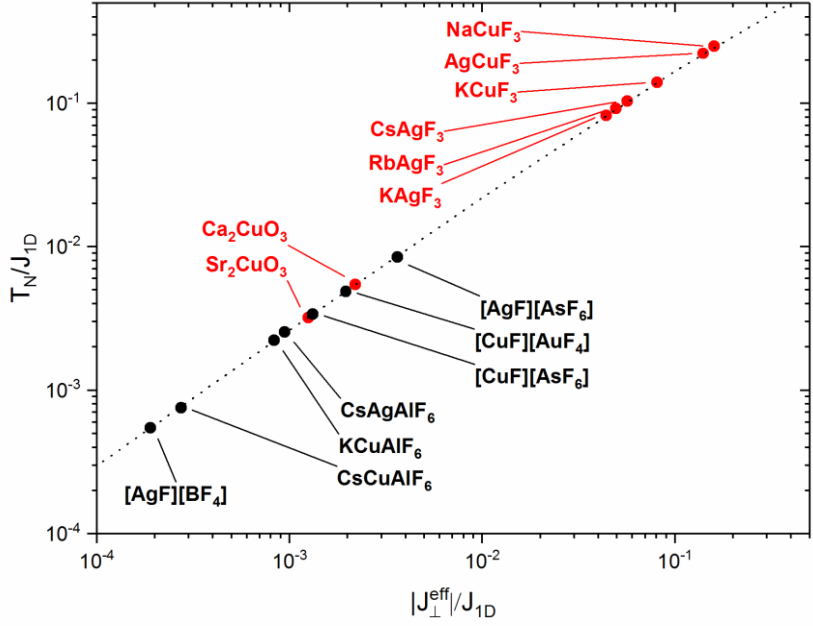

FIG. 8 The calculated dependence of the Néel temperature scaled by $J_{1D}$ ($T_N/J_{1D}$) as a function of $|J_\perp^{eff}|/J_{1D}$. Systems with $A_{1g}$ chains are marked in black, those exhibiting $B_{1g}$ chains in red. The dotted line represents the dependence given by Eq. (4).

For the studied compounds we find that systems exhibiting $A_{1g}$ chains are generally characterized by larger magnetic anisotropies (*i.e.* lower $|J_\perp^{eff}|/J_{1D}$ ratio) than $B_{1g}$ systems, as evident from Fig. 8. There are several reasons behind this. One is that chains built of compressed $Cu^{2+}/Ag^{2+}$ octahedra lead to stronger AFM intra-chain interactions than those seen for elongated octahedra.[8] Secondly most of the $A_{1g}$ systems studied here have $M^{2+}$ concentrations smaller than corresponding $B_{1g}$ compounds, and this greater dilution of spin sites leads to weaker inter-chain interactions. However we find that for [AgF][BF] the magnetic anisotropy remains high even upon application of high-pressure[73,74] (up to 100 kbar).[68]

## IV. CONCLUSIONS

We performed hybrid density functional calculations of magnetic coupling constants for a variety of 1D AFM systems. Our calculations reproduce with good accuracy experimental $T_N/J_{1D}$ ratio. Importantly we find that fluoride systems containing $Ag^{2+}$ and $Cu^{2+}$ cations in compressed octahedral coordination should constitute a novel family of quasi-1D systems exhibiting strong intra-chain AFM coupling and simultaneously very weak inter-chain coupling. For example, we show that for [AgF][BF$_4$] the computed values of $|J_\perp^{eff}|/J_{1D}$ and $T_N/J_{1D}$ are equal to $1.9 \cdot 10^{-4}$ and $5.5 \cdot 10^{-4}$, respectively; these values are smaller by ca. one order of magnitude than the calculated values for well researched $Sr_2CuO_3$. Hence, [AgF][BF$_4$] might be one of the best realizations of a quasi-1D AFM system. Importantly, this compound and other members of this group feature $d^9$ $Ag^{2+}$/$Cu^{2+}$ cations with one unpaired electron occupying the local $d(z^2)$ orbital rather than the $d(x^2–y^2)$ one (as found for $Sr_2CuO_3$). We hope that this study will motivate experimental investigation into the properties of this compound and its siblings, in particular in the context of novel phenomena that might arise from the different local symmetry of the spin-carrying orbital.

**V. ACKNOWLDEGEMENTS**

WG thanks the National Science Centre of the Republic of Poland (NCN) for HARMONIA grant (2012/06/M/ST5/00344) and ICM UW for time at OKEANOS supercomputer (ADVANCE Plus, GA67-13). Comments from Prof. José Lorenzana and Dr. M. Derzsi are greatly appreciated. This work is dedicated to Prof. Bogumił Jeziorski at his 70$^{th}$ birthday.

# SUPPLEMENTAL MATERIAL

# Large magnetic anisotropy in quasi one-dimensional spin-½ fluoride antiferromagnets with a d($z^2$)$^1$ ground state

D. Kurzydłowski,[1,2]* and W. Grochala[1]*

**CONTENTS**

# I. Structures of fluorides containing $A_{1g}$ chains

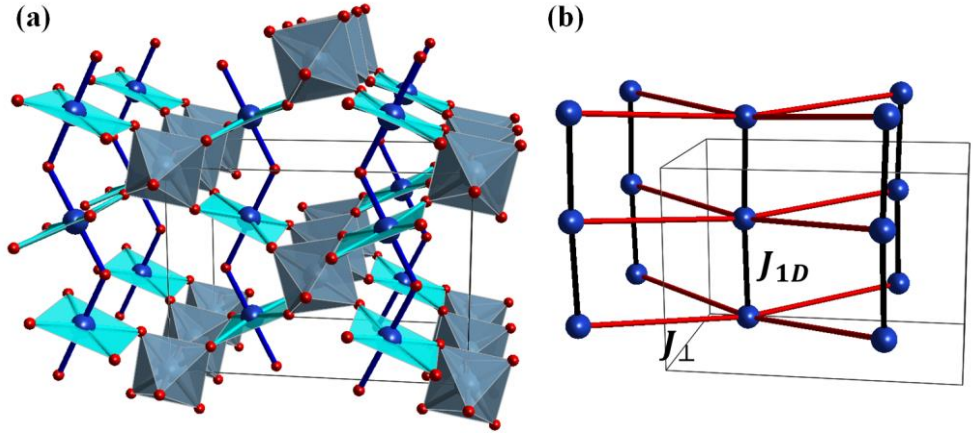

FIG. S 1 (a) Structure of $CsAgAlF_6$, $KCuAlF_6$, and $CsCuAlF_6$ (the three compounds are iso-structural); blue balls depict Al atoms, while K/Cs atoms were omitted for clarity; (b) Schematic description of the connectivity of the $J_{1D}$ (black) and $J_\perp$ (red) superexchange paths.

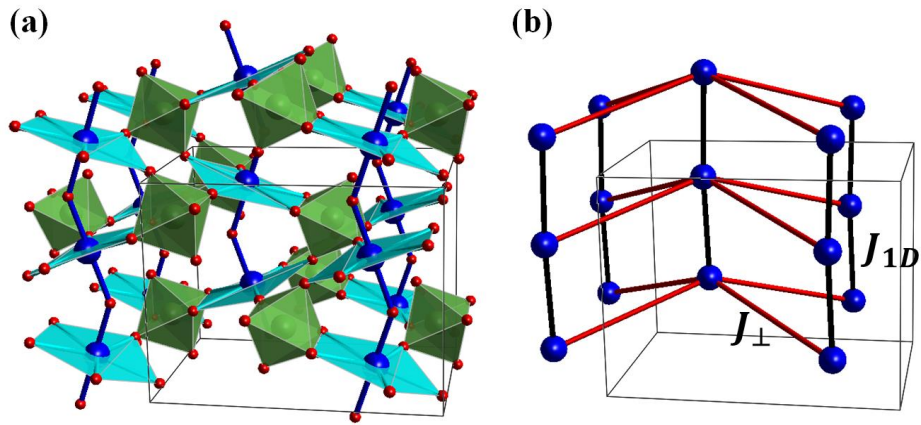

FIG. S 2 (a) Structure of $[AgF][AsF_6]$ (green polyhedral mark $AsF_6^-$ anions); (b) Schematic description of the connectivity of the $J_{1D}$ (black) and $J_\perp$ (red) superexchange paths.

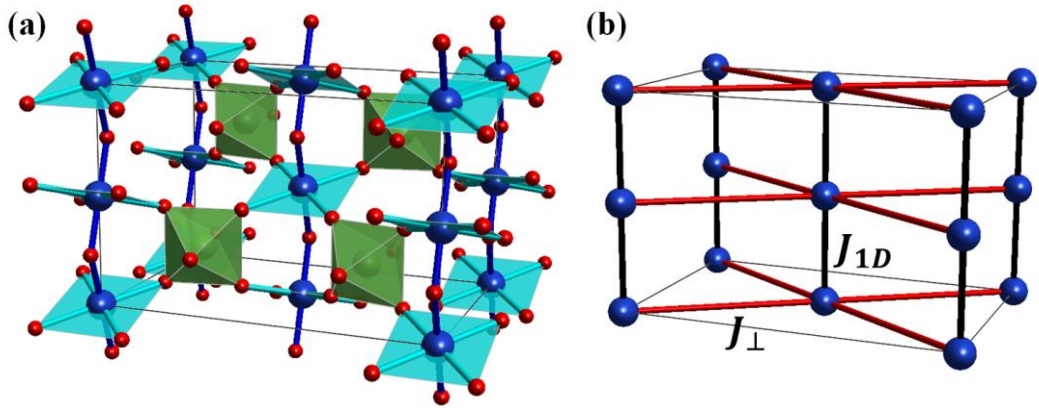

FIG. S 3 (a) Structure of $[CuF][AsF_6]$ (green polyhedral mark $AsF_6^-$ anions); (b) Schematic description of the connectivity of the $J_{1D}$ (black) and $J_\perp$ (red) superexchange paths.

## II. Magnetic states of [CuF][AuF$_4$]

TABLE S 1 Magnetic states of [CuF][AuF$_4$]. Sites with spin up/down are indicated with a +/− sign (their labelling follows that of the figure below.

| State | Site | | | | | | | | Energy per f.u. |
|---|---|---|---|---|---|---|---|---|---|
|  | 1 | 2 | 3 | 4 | 5 | 6 | 7 | 8 |  |
| F1 | + | + | + | + | + | + | + | + | $0.25J_{1D} + 0.25J_\perp^1 + 0.25J_\perp^2 + 0.25J_\perp^3 + E_{nm}$ |
| A1 | + | + | + | + | − | − | − | − | $-0.25J_{1D} + 0.25J_\perp^1 - 0.25J_\perp^2 + 0.25J_\perp^3 + E_{nm}$ |
| A2 | + | + | − | − | − | − | + | + | $-0.25J_{1D} + 0.25J_\perp^1 - 0.25J_\perp^2 - 0.25J_\perp^3 + E_{nm}$ |
| A3 | + | − | + | − | − | + | − | + | $-0.25J_{1D} - 0.25J_\perp^1 + 0.25J_\perp^2 + 0.25J_\perp^3 + E_{nm}$ |
| A4 | + | + | + | − | − | − | + | − | $-0.25J_\perp^2 + E_{nm}$ |

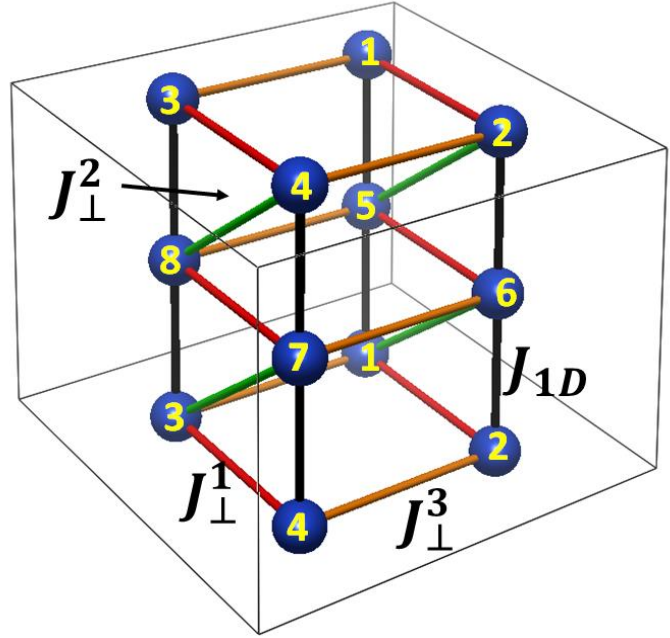

Equations used for the calculations of the magnetic coupling constants:

$J_{1D} = F1 - A1 - A2 - A3 - 2A4,$      (S1)

$J_\perp^1 = F1 + A1 + A2 - A3 - 2A4,$      (S2)

$J_\perp^2 = F1 - A1 + A2 + A3 - 2A4,$      (S3)

$J_\perp^3 = 2A1 - 2A2.$      (S4)

## III. Magnetic states of [AgF][BF$_4$]

TABLE S 2 Magnetic states of [AgF][BF$_4$] . Sites with spin up/down are indicated with a +/− sign (their labelling follows that of the figure below.

| State | Site | | | | | | | | Energy per f.u. |
|---|---|---|---|---|---|---|---|---|---|
| | 1 | 2 | 3 | 4 | 5 | 6 | 7 | 8 | |
| F1 | + | + | + | + | + | + | + | + | $0.25J_{1D} + 0.5J_\perp^1 + 0.5J_\perp^2 + 0.5J_\perp^3 + E_{nm}$ |
| A1 | + | + | + | + | − | − | − | − | $-0.25J_{1D} + 0.5J_\perp^1 - 0.5J_\perp^2 + 0.5J_\perp^3 + E_{nm}$ |
| A2 | + | + | − | − | − | − | + | + | $-0.25J_{1D} - 0.5J_\perp^1 + 0.5J_\perp^2 + 0.5J_\perp^3 + E_{nm}$ |
| A3 | + | + | − | − | + | + | − | − | $0.25J_{1D} - 0.5J_\perp^1 - 0.5J_\perp^2 + 0.5J_\perp^3 + E_{nm}$ |
| A4 | + | − | − | − | − | + | + | + | $-0.25J_{1D} + E_{nm}$ |

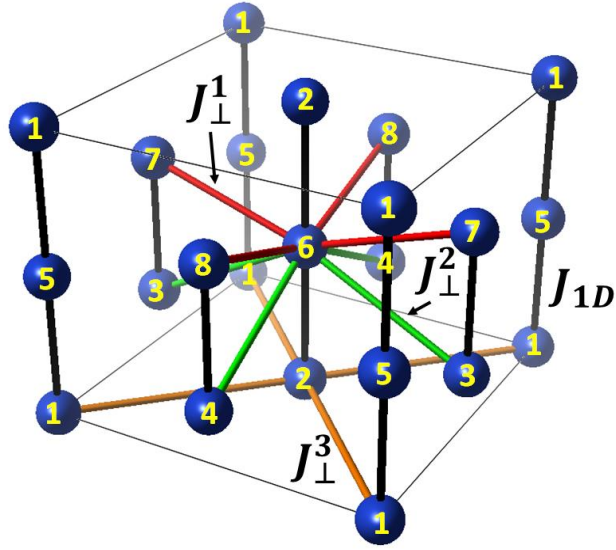

Equations used for the calculations of the magnetic coupling constants:

$J_{1D} = F1 - A1 - A2 + A3,$ (S5)

$J_\perp^1 = 1/2(F1 + A1 - A2 - A3),$ (S6)

$J_\perp^2 = 1/2(F1 - A1 + A2 - A3),$ (S7)

$J_\perp^3 = A1 + A2 - 2A4.$ (S8)

**IV. Pressure dependence of the magnetic anisotropy and the scaled Néel temperature for [AgF][BF$_4$]**

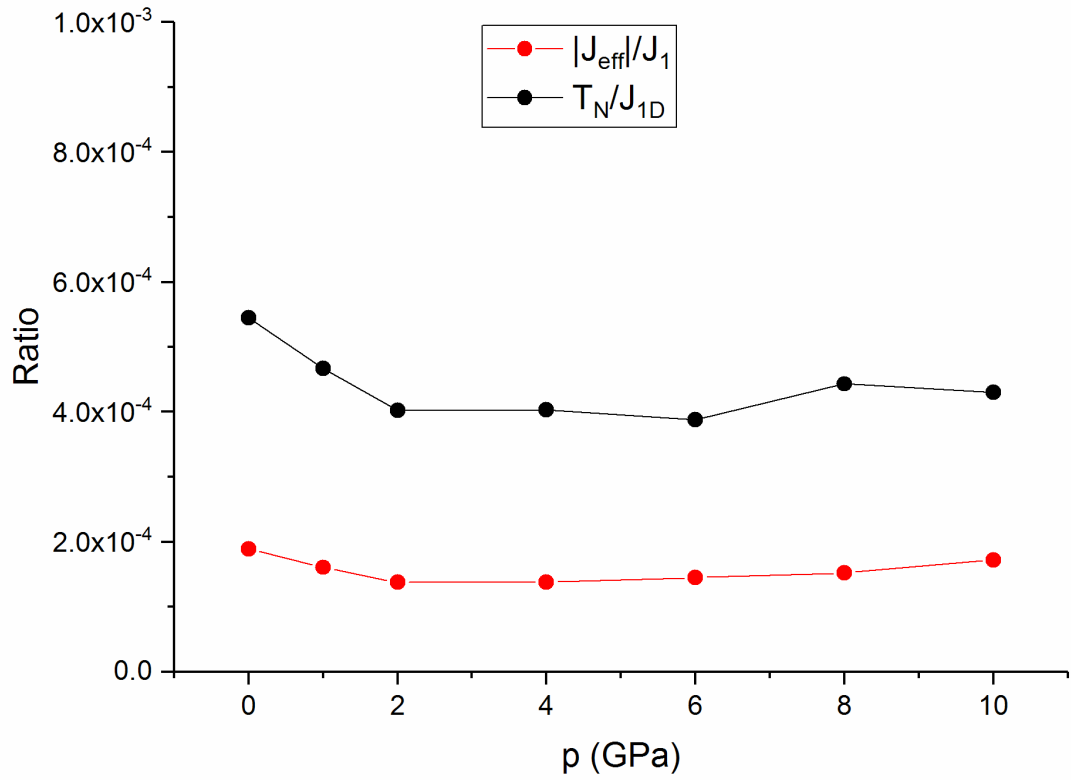